# Ultralong Wave Focusing via Generalized Luneburg Lens

Liuxian Zhao[1,2], Miao Yu[2,3,*]

[1]Institute of Sound and Vibration Research, Hefei University of Technology, 193 Tunxi Road, Hefei 230009, China

[2]Institute for Systems Research, University of Maryland, College Park, MD, 20742, USA

[3]Department of Mechanical Engineering, University of Maryland, College Park, Maryland 20742, USA

*Author to whom correspondence should be addressed: mmyu@umd.edu

**ABSTRACT**

In this paper, a novel gradient index (GRIN) structural lens based on the concept of generalized Luneburg lens (GLL) is proposed. This lens allows for the realization of double foci and localization of energy flow between the two focal spots, thereby achieving ultralong focusing. The double-foci GRIN lens consists of two concentric circular regions with varying thickness defined in a thin plate structure. The two concentric circular regions are designed to realize continuous change of refractive indices with different profiles. Numerical simulations and experimental studies are performed to obtain the maximum displacement amplitude, full length at half maximum (FLHM), and full width at half maximum (FWHM) of the focal region of the lens. The results demonstrate that ultralong subwavelength focusing can be achieved for a broadband frequency range. In addition, our results show that the FLHM and FWHM can be tailored through the design of the focal length of the GLL. This offers a simple and flexible approach of engineering the GLL focusing characteristics and energy distributions for many applications.

# 1. Introduction

Gradient index (GRIN) lens has attracted much attention in recent years for operation and control of optical and acoustic wave propagation [1-4]. A GRIN lens entails a spatial variation of refractive index, which can be achieved by locally changing the geometric parameters of subwavelength unit cells. The effective refractive index of a GRIN lens depends on the filling ratio of unit cell scatterers [5, 6]. The first GRIN lens for structural waves was proposed by Lin *et al.*, which demonstrated the manipulation of the structural wave propagation along the desired trajectories by tailoring the filling ratio of unit cells [7]. For structural wave propagation in thin plates, the effective velocity of structural wave was shown to be directly related to the plate thickness, therefore, the GRIN lens can be achieved through varying the plate thickness [8-11].

One of the most interesting applications of structural wave propagation in plates is the vibration-based energy harvesting [12-15]. Focusing of plane structural waves with GRIN lens has been investigated for achieving enhanced energy harvesting. As environmental vibrations are usually in the far field, these excitations can be approximated as plane wave excitations and the energy can be collected at the focusing location of the lens. For example, Tol *et al.* [16] proposed a GRIN lens consisting of cylinder arrays with changing height and demonstrated that a piezoelectric transducer placed at the lens focus could produce a higher power of up to three times. Zhao *et al.* [17] designed a GRIN structural Luneburg lens based on continuous variation of plate thickness and demonstrated energy harvesting enhancement of nine times at the focal location of the lens. In another study, Zhao *et al.* [18] investigated a planar GRIN lens with a varying thickness, which achieved an amplitude of displacement at the focus about 30~40 times of that of incident wave. It should be noted that in all of the above-mentioned techniques the mechanical vibration energy was focused on a small single focus having a short full length at half maximum (FLHM).

Based on the principles of Luneburg lens [19-22], Sochacki [23] proposed a multi-foci optical GRIN lens. By changing refractive index profiles, the focal point can be tuned to locate either inside or outside of the lens and multiple focal points can be generated. Specifically, a double-foci generalized Luneburg lens (GLL) enables the positions of two focal points to be tailorable and allow energy localization between the two focal spots. Double-foci GLLs have been investigated in several recent studies. For example, Mao *et al.* [24] explored a tunable nanojet obtained with generalized optical Luneburg lens for achieving ultralong subwavelength focusing between the two foci. Chou *et al.* [25] proposed a double-foci GLL working in the microwave regime based on a metasurface consisting of circular metallic patches fabricated on a grounded dielectric substrate.

In this work, inspired by the concept of GLL in the electromagnetic regime, we explore a broadband double foci GLL for structural wave manipulation, which enables the realization of an ultralong focal region between the two foci. The refractive index variation in the structural GLL is realized by changing the lens thickness in order to tune the velocity of structural waves.

## 2. Generalized Luneburg Lens Design

The principle of the proposed structural GLL is illustrated in Figure 1(a), the lens is composed of two concentric circles with outer radius of *R* and inner radius of *R'*. As a plane wave interacts with the structural GLL, double foci can be produced. All the rays passing through the inner circle are focused with a focal length $F_1$, and all the rays passing through the outer circle are focused with a focal length $F_2$. The configuration of rays in such a medium can be described as follows, with the detailed derivations provided in [23]:

$$\begin{cases} \int_r^{R'} \frac{k}{r\sqrt{s^2-k^2}} dr + \int_{R'}^{R} \frac{k}{r\sqrt{s^2-k^2}} dr = \frac{1}{2}\arcsin\left(\frac{k}{F_1}\right) + \arccos(k), & 0 \leq s \leq P_a \\ \int_r^{R} \frac{k}{r\sqrt{s^2-k^2}} dr = \frac{1}{2}\arcsin\left(\frac{k}{F_2}\right) + \arcsin(k), & P_a \leq s \leq R \end{cases} \quad (1)$$

where $r$ is radial distance, $s = rn/R$, $\varphi$ is the angle between the radius distance $r$ and the tangential line, and $k = n(r)r \cdot \sin(\varphi)$. Note that $P_a$ is parameter within the range of 0 to $R$. The inner circle radius can be calculated based on the value of $P_a$; that is, $P_a = R'n(R')$, where $n(R')$ is the refractive index of the inner circle. Based on Equation (1), the refractive index profile is obtained as:

$$n(s) = \begin{cases} e^{\{\omega_2(s,F_1,P_a)-\omega_2(s,F_2,P_a)+\omega_1(s,F_2)\}}, & 0 \leq s < P_a \\ e^{\omega_1(s,F_2)}, & P_a \leq s \leq R \end{cases} \quad (2)$$

where $\omega_1(s,F_2) = \frac{1}{\pi}\int_s^1 \frac{\sin^{-1}(k/F_2)}{\sqrt{k^2-s^2}}dk$ and $\omega_2(s,F_i,P_a) = \frac{1}{\pi}\int_s^{P_a} \frac{\sin^{-1}(k/F_i)}{\sqrt{k^2-s^2}}dk$.

When flexural waves propagate through a thin plate structure with a variable thickness, the phase velocity $c_p$ is a function of the plate thickness $h$, which can be expressed as $c_p = (\frac{E\omega^2 h^2}{12(1-\nu^2)\rho})^{\frac{1}{4}}$. Here, $\rho$ is the material density, $\nu$ is the Poisson's ratio, $E$ is the material Young's modulus, and $\omega$ is the natural frequency. According to the Snell's law, $n = c_0/c_p$, where $c_0$ is the wave speed of the input wave on a constant structural thickness ($h_0$). The profile of $n$ can be derived as a function of the plate thickness [17]: $n = \sqrt{h_0/h}$. Therefore, the variable thickness profile of the double-foci GLL defined on a thin plate structure is provided as:

$$h(s) = \frac{h_0}{n^2} = \begin{cases} e^{2\{\omega_2(s,F_1,P_a)-\omega_2(s,F_2,P_a)+\omega_1(s,F_2)\}}h_0, & 0 \leq s < P_a \\ e^{2\omega_1(s,F_2)}h_0, & P_a \leq s \leq R \end{cases} \quad (3)$$

In our numerical and experimental studies, we chose the following representative geometric parameters: $h_0 = 4$ mm, $R = 100$ mm, and $F_1 = 1.2R$. Different focal lengths $F_2$ ($F_2 = 1.6R$ and $F_2 = 2.5R$) were considered. Based on Equations (2) and (3), the distribution of refractive indices along the radial distance is obtained, as shown in Figure 1(b).

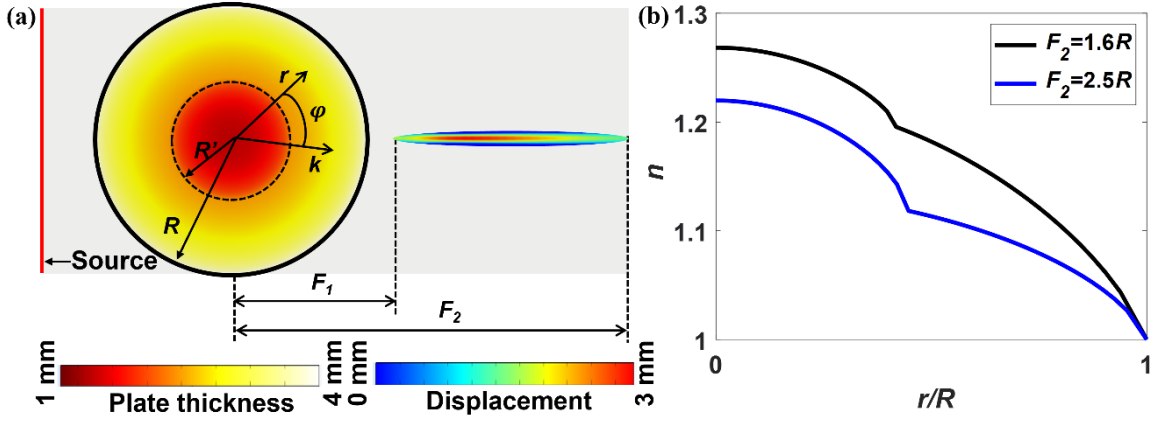

**Figure 1:** Mechanism of the double-foci GLL and its design principle. (a) Schematic of double-foci GLL for flexural wave manipulation. (b) Distribution of refractive index along the radial distance for different focal length $F_2$. The input displacement amplitude of the source is $I_0=1$ mm.

## 3. Numerical and Experimental Studies

The proposed GLL can be analysed by using different numerical methods, such as the spectral method [26]. In this study, we performed numerical simulations based on the finite element method to investigate the proposed GLL for flexural wave focusing on a thin aluminium plate. The focal length $F_1$ was kept as a constant value of $F_1=1.2R$. In the numerical simulations, the following material properties were used for the plate: the Young's modulus of 70 GPa, density of 2700 kg/m$^3$, and Poisson's ratio of 0.33. The plate dimensions of 1000 mm × 400 mm × 4 mm were used. Both time and frequency domain simulations were performed by using COMSOL software. Absorbing boundary conditions (PML) were utilized to reduce reflections from the plate edges. In order to validate the performance of structural Luneburg lens for broadband ultralong subwavelength focusing without the effect of structural damping, there is no damping added to the structure in the simulations. The excitation signal was a harmonic signal with a central frequency of 100 kHz, which was located at a distance of 0.18 m from the lens center. In order to guarantee the convergence of simulations for frequencies up to 200 kHz, we used a maximum element size of 0.5 mm. This condition ensures that there are at least 20 elements per wavelength at the frequency of 200 kHz [27].

The performance of double-foci GLL was also studied experimentally. The lens was fabricated on an aluminium plate (6061 aluminium from McMaster-Carr) with dimensions 650 mm × 650 mm × 4 mm. The fabricated GLL and the experimental arrangement are shown in Figure 2. Absorbing clays were applied to the plate boundaries to reduce reflections. Ten piezoelectric transducers with dimensions of 20 mm× 15 mm × 1 mm (SMPL20W15T1R111, STEINER & MARTINS, INC.) were attached to the plate to excite plane waves. Adhesive (FastCap 2P-10 Super Glue) was used to bond the piezoelectric transducers to the structure. Two GRIN lens with different focal lengths of $F_2= 1.6R$ and $F_2= 2.5R$ were fabricated (see Figures 2 (b)-(c)). In this experiment, the two short edges of the plate were constrained. The structural vibrations were measured by using a laser vibrometer (Polytec PSV-400) with a scanning area of 600 mm x 400 mm (see Figures 2 (b)-(d)). The performance of double-foci GRIN lens was investigated in the time domain.

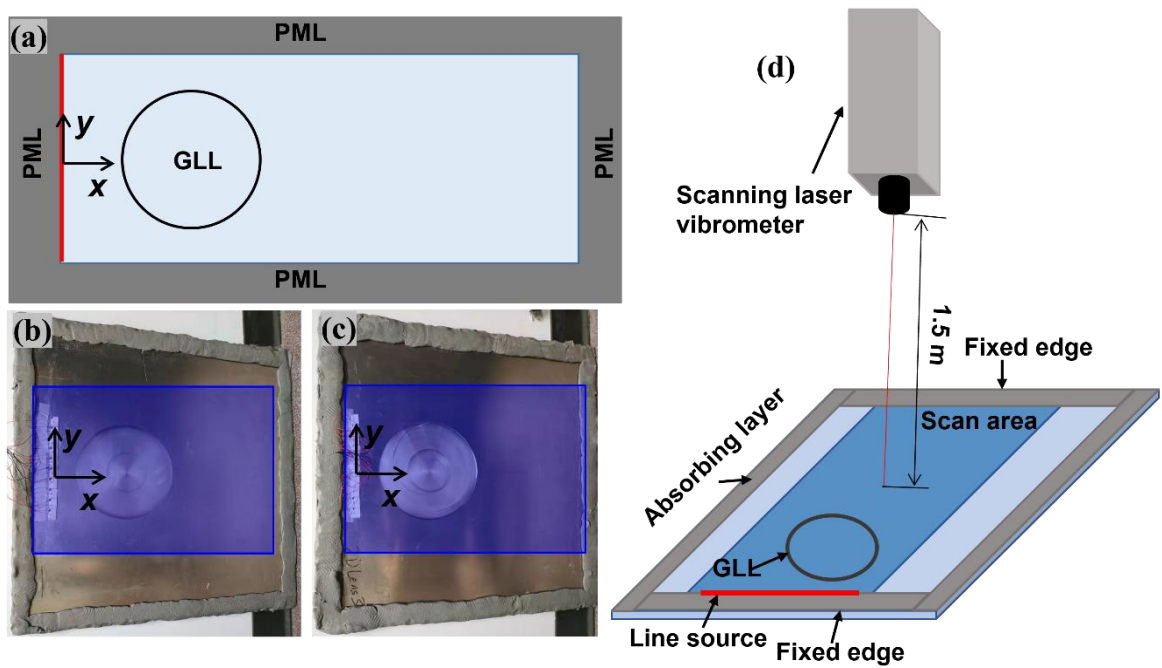

**Figure 2: Numerical model and experimental setup. (a) Numerical model used to simulate the structural wave propagation. The red line represents the input source. (b) and (c) Fabricated GRIN lenses with focal lengths of $F_2=1.6R$ and $F_2=2.5R$, respectively. The blue areas are used for measurements (with dimensions of 600 mm x 400 mm). (d) Schematic of the experimental setup. A scanning laser vibrometer was used to measure the full-field wave propagations in the scan area. All four plate edges were covered with clay and two edges were constrained.**

First, finite element simulations were conducted to validate the performance of subwavelength double-foci GLL at different frequencies and different focal lengths ($F_1=1.2R$ and $F_2=2.5R$). A plane wave was excited at $x = 0$ m (at a distance of 0.18 m from the lens center) at four different frequencies of 50 kHz, 100 kHz, 150 kHz, and 200 kHz. The obtained normalized waveforms and amplitude fields are provided in Figures 3 (a)-(b). These results demonstrate that an ultralong focus can be achieved at different frequencies.

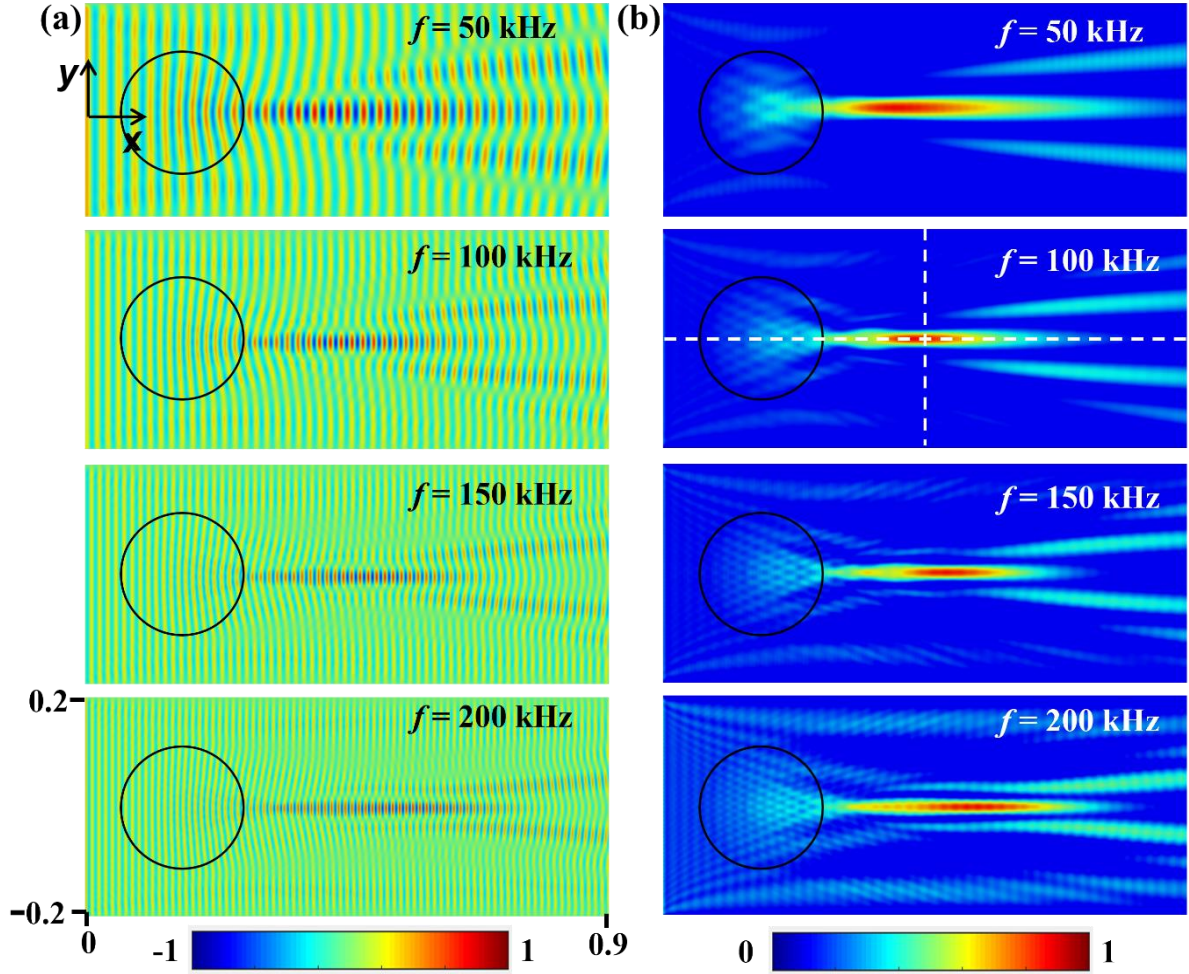

**Figure 3: Numerical simulations of the steady state response of wave propagation through the double foci GLL at different frequencies. (a) Normalized waveform and (b) displacement amplitude field distributions. The black circle outlines the structural lens, and the colour indicates the normalized displacement field, which is normalized to the maximum value of the displacement amplitude field. The unit of the length scale of axes is meter.**

Next, to characterize the ultralong subwavelength focusing performance of the structural GLL, the displacement amplitude field distributions of the double-foci GLL along the *x* axis through the lens center (horizontal white dotted line in Figure 3(b)) and the *y* axis through the peak value of amplitude (vertical white dotted line in Figure 3(b)) for different focal lengths $F_2$ (1.6$R$, 1.9$R$, 2.2$R$, and 2.5$R$) were obtained at 100 kHz (***I*$_1$** and ***I*$_2$**, respectively). The output/input displacement ratios (***I*$_1$**/***I*$_0$** and ***I*$_2$**/***I*$_0$**) were plotted in Figures 4 (a) and (b), respectively. In addition, the full length at half maximum (FLHM) and full width at half maximum (FWHM) of the ultralong focusing were obtained based on the amplitude along *x* and *y* directions, and the results are shown in Figures 4(c) and (d), respectively. Here, the FLHM is defined as the length where the displacement is over half of its maximum value along the *x* direction, starting from *x*=0 m. The FWHM is the width between two locations along the *y* direction where the displacements are above half of its maximum value. It can be seen that the FLHM increases as the focal length $F_2$ goes up. At the focal length $F_2$ =2.5$R$, the FLHM can be up to 15 . For all the simulated focal lengths $F_2$, the obtained FWHM are smaller than $\lambda$. These results demonstrate that the GLL can produce a subwavelength scale focus. Note that in this study the input displacement amplitude of 1 mm is used for proof of concept. The results (FLHM and FWHM) will not change if a smaller displacement amplitude value is used.

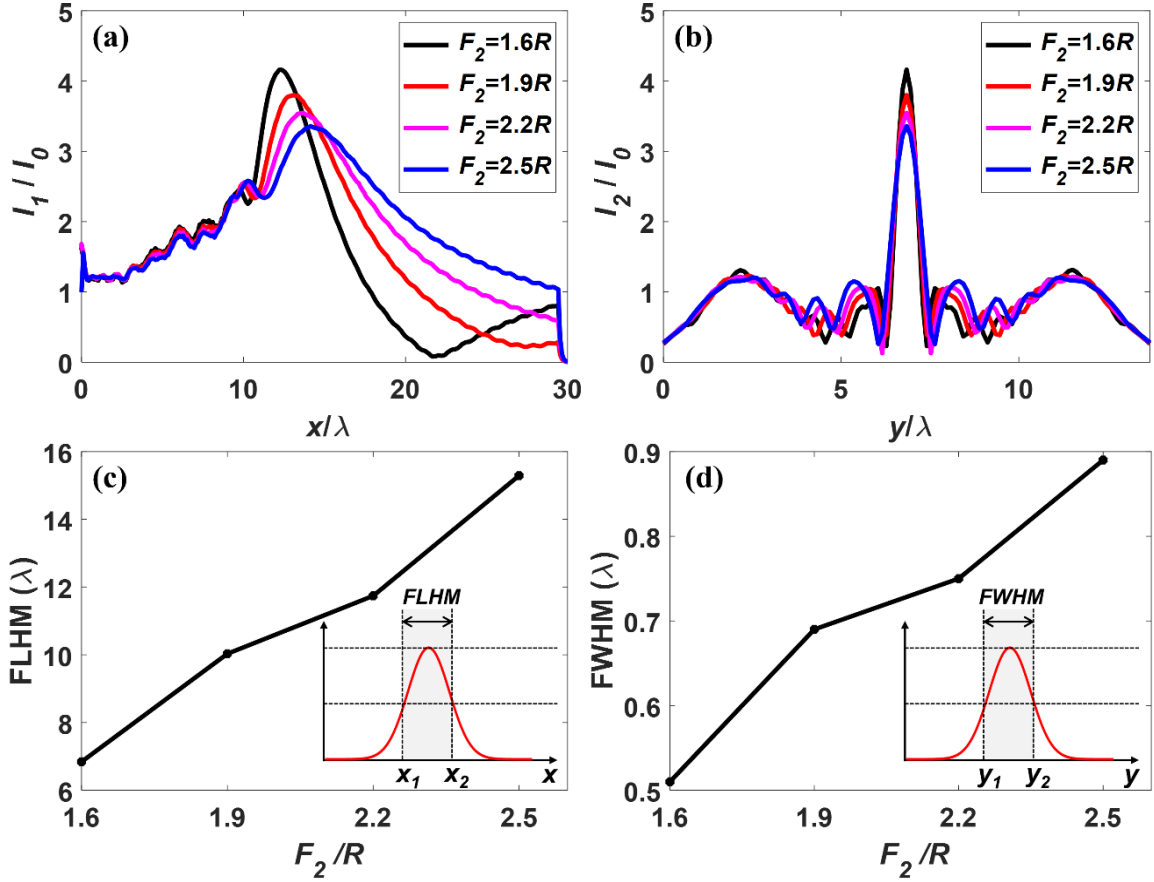

**Figure 4:** Numerical simulations of the output/input displacement ratio of the double-foci GLL along the *x* axis (a) and the *y* axis at maximum amplitude spots (b) obtained for different focal length $F_2$. (c) and (d) FLHM and FWHM obtained for different focal length $F_2$. The insets in (c) and (d) illustrate the definitions of FLHM and FWHM, respectively.

Furthermore, in order to show the double focusing properties of the structural GLL at each time instants, transient analyses were conducted numerically and experimentally with different values of $F_2$ ($F_2$=1.6R and $F_2$=2.5R). A line source along the *y* direction was applied to generate a plane wave. The excitation signal was a 3-count tone burst signal with a central frequency of 100 kHz and a 3 dB bandwidth of 67 kHz (from 66 kHz to 133 kHz). The obtained waveforms for both $F_2$=1.6R and $F_2$=2.5R at time instants of $t$ = 0.04, 0.11, 0.16, and 0.21 ms are provided in Figures 5(a)-(b) for numerical simulations and Figures 5 (c)-(d) for experimental measurements.

At the time instant of 0.04 ms, the generated plane waves propagated forward. At 0.11 ms, the waves propagated within the GRIN lens and the wavefront became curved. At 0.16 ms, the planes waves became narrower and focused for both $F_2=1.6R$ and $F_2=2.5R$. At 0.21 ms, the focused waves became divergent for $F_2=1.6R$, while the waves remained to be focused for $F_2=2.5R$. These results are consistent with those shown in Figure 4, where the FLHM obtained for $F_2=2.5R$ is longer than that obtained for $F_2=1.6R$. The experimental measurements agree well with the numerical simulation results.

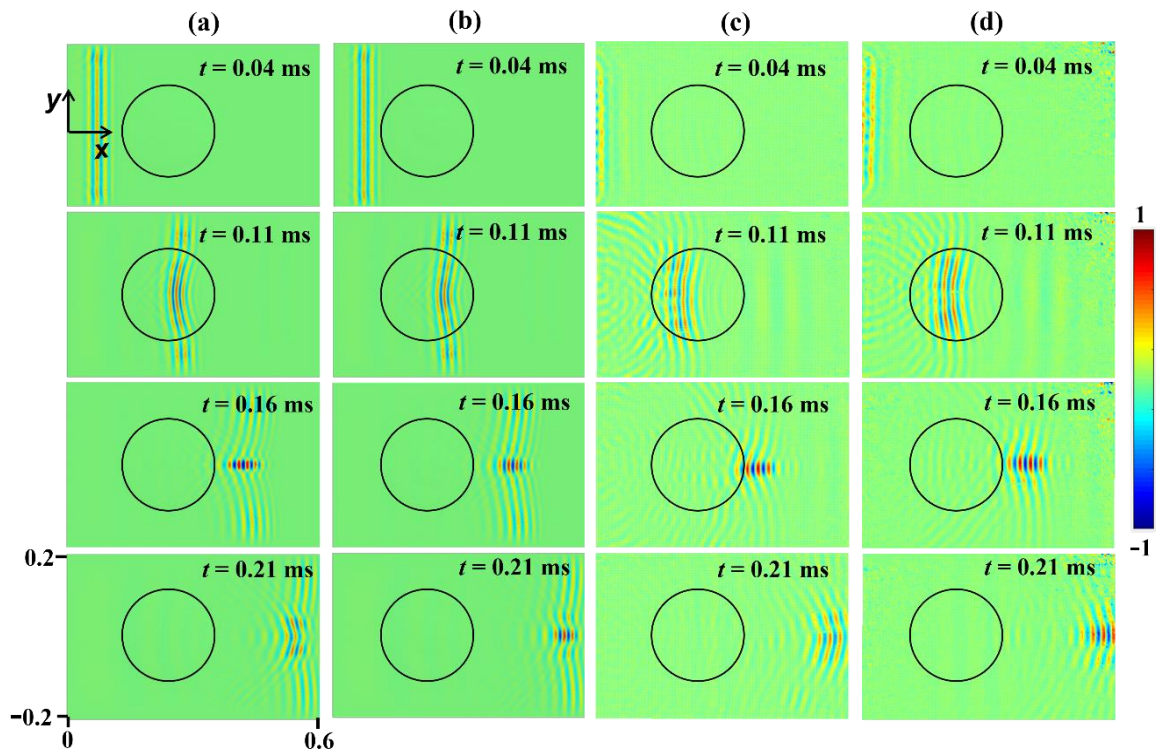

**Figure 5: Transient response of structural GLL for flexural wave double focusing at time instants of $t = 0.04$ ms, $t = 0.11$ ms, $t = 0.16$ ms, and $t = 0.21$ ms. (a) and (b) are numerical simulations obtained for GLL with a focal length of $F_2=1.6R$ and $F_2=2.5R$, respectively. (c) and (d) are experimental measurements obtained for GLL with a focal length of**

**$F_2$=1.6R and $F_2$=2.5R, respectively. The black circle outlines of the structural GLL, and the colour indicates the normalized displacement field. The unit of the axis scale is meter. The amplitude of the waveform is normalized to its maximum value.**

In addition, to investigate the performance of double-foci structural GLL, the transient response at the two different focal lengths is provided in Figures 6 in terms of time–space data obtained through both numerical simulations and experimental measurements at the frequency of 100 kHz. For both selected focal lengths, it can be clearly seen that ultralong focusing can be achieved when the flexural waves propagate through the GLL. The difference between the two cases is the length of the focusing region in terms of FLHM. In experiment, the FLHM reaches 10$\lambda$ for $F_2$=2.5R compared to that of 6$\lambda$ for $F_2$=1.6R. The experimental results are in good agreement with the numerical simulation results, which clearly validate the ultralong focusing capability of the proposed structural GLL.

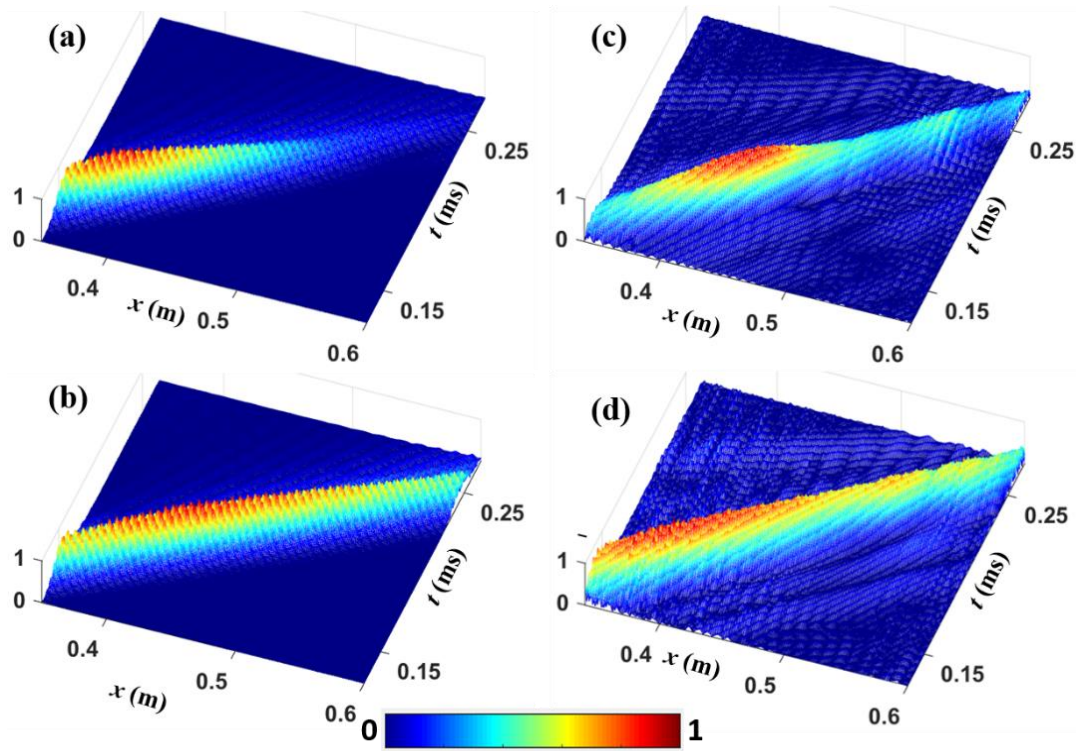

**Figure 6: Numerical simulations and experimental measurements of the GLL performance in time-space domain. (a) and (b) are the normalized displacement fields obtained in numerical simulations with focal length of $F_2$=1.6R and $F_2$=2.5R, respectively. (c) and (d) are the normalized displacement fields measured in experiments with focal length of $F_2$=1.6R and $F_2$=2.5R, respectively. The colorbars indicate the normalized displacement values shown in the $z$ axis.**

## 4. Conclusions

In conclusion, a novel design of gradient refraction index (GRIN) structural lens based on the generalized Luneburg lens (GLLs) was investigated, The lens was achieved by using a variable thickness concentric circular structure. This simple lens design renders a continuous gradient index, enabling broadband, ultralong subwavelength focusing of flexural wave. Numerical and experimental studies were carried out to explore the ultralong focusing properties of the proposed lens. The results show that the GRIN lens can achieve a large FLHM up to 15$\lambda$ at the focusing region, while the FWHM is limited to a subwavelength scale for a broadband frequency range. This work is expected to benefit vibration-based energy harvesting techniques for many applications.

**Conflict of Interest**

The authors declare no conflict of interest.

**Data Access Statement**

The data that support the findings of this study are available from the corresponding author upon reasonable request.

**Reference**

1. Tol, S., F.L. Degertekin, and A. Erturk, *Gradient-index phononic crystal lens-based enhancement of elastic wave energy harvesting.* Applied Physics Letters, 2016. **109**(6): p. 063902.
2. Liu, W., et al., *Manipulating light trace in a gradient-refractive-index medium: a Lagrangian optics method.* Optics Express, 2019. **27**(4): p. 4714-4726.
3. Jin, Y., B. Djafari-Rouhani, and D. Torrent, *Gradient index phononic crystals and metamaterials.* Nanophotonics, 2019. **8**: p. 685-701.
4. Park, J., et al., *Double-Focusing Gradient-Index Lens with Elastic Bragg Mirror for Highly Efficient Energy Harvesting.* Nanomaterials, 2022. **12**(6).
5. Titovich, A.S., A.N. Norris, and M.R. Haberman, *A high transmission broadband gradient index lens using elastic shell acoustic metamaterial elements.* The Journal of the Acoustical Society of America, 2016. **139**(6): p. 3357-3364.
6. Ma, F., et al., *An underwater planar lens for broadband acoustic concentrator.* Applied Physics Letters, 2022. **120**(12): p. 121701.
7. Lin, S.-C.S., et al., *Gradient-index phononic crystals.* Physical Review B, 2009. **79**(9): p. 094302.
8. Climente, A., D. Torrent, and J. Sánchez-Dehesa, *Gradient index lenses for flexural waves based on thickness variations.* Applied Physics Letters, 2014. **105**(6): p. 064101.
9. Jin, Y., et al., *Gradient Index Devices for the Full Control of Elastic Waves in Plates.* Scientific Reports, 2016. **6**(1): p. 24437.
10. Jin, Y., et al., *Simultaneous control of the S0 and A0 Lamb modes by graded phononic crystal plates.* Journal of Applied Physics, 2015. **117**(24): p. 244904.
11. Zhao, L., S.C. Conlon, and F. Semperlotti, *Broadband energy harvesting using acoustic black hole structural tailoring.* Smart Materials and Structures, 2014. **23**(6): p. 065021.
12. Wang, J., et al., *Perspectives in flow-induced vibration energy harvesting.* Applied Physics Letters, 2021. **119**(10): p. 100502.
13. Aridogan, U., I. Basdogan, and A. Erturk, *Random vibration energy harvesting on thin plates using multiple piezopatches.* Journal of Intelligent Material Systems and Structures, 2016. **27**(20): p. 2744-2756.
14. Liang, H., G. Hao, and O.Z. Olszewski, *A review on vibration-based piezoelectric energy harvesting from the aspect of compliant mechanisms.* Sensors and Actuators A: Physical, 2021. **331**: p. 112743.
15. Hegendörfer, A., P. Steinmann, and J. Mergheim, *Nonlinear finite element system simulation of piezoelectric vibration-based energy harvesters.* Journal of Intelligent Material Systems and Structures, 2021. **33**(10): p. 1292-1307.
16. Tol, S., F.L. Degertekin, and A. Erturk, *3D-printed phononic crystal lens for elastic wave focusing and energy harvesting.* Additive Manufacturing, 2019. **29**: p. 100780.
17. Zhao, L., C. Lai, and M. Yu, *Modified structural Luneburg lens for broadband focusing and collimation.* Mechanical Systems and Signal Processing, 2020. **144**: p. 106868.
18. Zhao, J., et al., *Broadband sub-diffraction and ultra-high energy density focusing of elastic waves in planar gradient-index lenses.* Journal of the Mechanics and Physics of Solids, 2021. **150**: p. 104357.
19. Luneburg, R.K., *Mathematical Theory of Optics*. 1944, Berkeley, CA: University of California Press.
20. Zhao, L., T. Horiuchi, and M. Yu, *Acoustic waveguide based on cascaded Luneburg lens.* JASA Express Letters, 2022. **2**(2): p. 024002.
21. Zhao, L., T. Horiuchi, and M. Yu, *Broadband acoustic collimation and focusing using reduced aberration acoustic Luneburg lens.* Journal of Applied Physics, 2021. **130**(21): p. 214901.
22. Zhao, L., T. Horiuchi, and M. Yu, *Broadband ultra-long acoustic jet based on double-foci Luneburg lens.* JASA Express Letters, 2021. **1**(11): p. 114001.
23. Sochacki, J., *Multiple-foci Luneburg lenses.* Applied Optics, 1984. **23**(23): p. 4444-4449.
15